\documentclass[aip,jcp,twocolumn, floatfix]{revtex4}

\usepackage{graphicx}
\graphicspath{ {./images/} }
\usepackage{subfig}
\usepackage{xcolor}

\usepackage[colorlinks=true, linkcolor=black, citecolor=black, urlcolor=blue]{hyperref}
\usepackage{amssymb}

\usepackage{xcolor}

\usepackage{amsmath}
\usepackage{soul}

\usepackage{placeins} 
\usepackage{float}
\usepackage{soul}
\usepackage{xcolor}

\begin{document}

\title{The Ion-Activated Attractive Patchy Particle Model and Its Application to the Liquid-Vapour Phase Transitions}

\author{Furio Surfaro \thanks{Corresponding Author}}
\email{furio.surfaro@uni-tuebingen.de}
\author{Fajun Zhang}
\author{Frank Schreiber}
\affiliation{Institute of Applied Physics, University of T{\"u}bingen, 72076 T{\"u}bingen, Germany}
\author{Roland Roth}
\email{roland.roth@uni-tuebingen.de}
\affiliation{Institute of Theoretical Physics, University of T{\"u}bingen, 72076 T{\"u}bingen, Germany}

\begin{abstract}
Patchy particles are an intriguing subject of study and indeed a model system in the field of soft matter physics. In recent years, patchy particle models have been applied to describe a wide variety of systems, including colloidal crystals, macromolecular interactions, liquid crystals, and nanoparticle assemblies. Given the importance of the topic, rationalizing and capturing the basic features of these models is crucial to their correct application in specific systems. In this study, we extend the ion-activated attractive patchy particles model previously employed to elucidate the phase behavior of protein solutions in the presence of trivalent salts. Our extension incorporates the effect of repulsion between unoccupied and occupied binding sites, depicted as patches. Furthermore, we examine the influence of model parameters on the liquid-vapor coexistence region within the phase diagram, employing numerical methods. A deeper understanding of this model will facilitate a better comprehension of the effects observed in experiments.
\end{abstract}

\maketitle

\section{Introduction}
Patchy particles are a class of complex colloidal particles with anisotropic surface interactions that enables them to selectively interact and bind with each other in a specific orientation. In recent years the interest in patchy particles models is increasing in materials science, physics, chemistry, and biology due to their potential applications in self-assembly, drug delivery, catalysis, and many other fields.\cite{pawar2010fabrication,smallenburg2013patchy,russo2011re,heidenreich2020designer}The flexibility of patchy particle models allows for a wide range of possible interactions and structures to be studied.\cite{bianchi2006phase,de2006dynamics,feng2013dna,mcmanus2016physics,teixeira2017phase, bianchi2017inverse} These models can incorporate both directional and isotropic interactions between particles, as well as steric effects, to accurately capture the behaviour of complex systems, such as protein-salt solutions.\cite{petukhov2017entropic,haghmoradi2020beyond, sato2021clusters,fusco2013crystallization,kastelic2018theory,kastelic2015protein,matsarskaia2020multivalent} Here, we present an extension of our ion-activated patchy particle model that we used before to capture the phase behaviour of protein-salt mixtures.\cite{roosen2014ion,fries2017multivalent,fries2020enhanced} The aim of this work is to understand the effect of different parameters included in the model on the effective interactions and on the behaviour of the liquid-vapour phase transition. We use the Wertheim theory, which provides a theoretical framework for predicting the thermodynamics properties of the patchy particle systems.\cite{wertheim1984fluids,wertheim1984fluidsII,wertheim1986fluids,lindquist2016formation,braz2021phase} We will show with our approach that the anisotropic behaviour of the system, arises from the different population of particles with occupied and unoccupied patches upon ion binding to the surface. In this regard, our multicomponent system composed of particles with different occupied and unoccupied binding sites, is transformed and investigated as an effective single component system.
\section{Theory}
\subsection{Ion-Activated Patchy Particles Model}
In this section we focus on the basic theory of the ion-activated patchy particles model \cite{roosen2014ion}. In our patchy particles system we assume that the probability to have an occupied patch, that is an ion bound to a binding site, as a function of the salt or ions concentration $\Theta({c^r_s})$ is given by a Fermi-like distribution in the grand canonical ensemble (GCE): \cite{roosen2014ion}
\begin{equation} \label{eq:binding}
    \Theta = \frac{1}{1+\exp\big(\beta(\epsilon_b -\mu_s(c^{r}_s))\big)},
\end{equation}
where $\beta=1/(k_B T)$ is the inverse temperature, $\epsilon_b$ is the binding energy between a salt ion and a patch on the particle surface. We assume that the patches are independent and possess the same energy $\epsilon_b$, which is kept constant and independent of the salt concentration. $\mu_s$ is the chemical potential of the salt in the reservoir, that can be approximated by the ideal gas expression $\mu^{r}_s(c^{r}_s)\approx k_B T \ln(c^{r}_s/\rho_0)$, where $\rho_0$ is the density of the reference state.

In our model with the assumption of $m$ independent patches per particle, the probability of finding $i$ patches occupied by ions is given by the binding probability $\Theta$ of a single patch via a binomial distribution \cite{roosen2014ion}
\begin{equation} \label{eq:prob}
    p(m, i) = \Theta^{i} (1-\Theta)^{m-i} \binom{m}{i},
\end{equation}
where $q= (1-\Theta)$ is the non binding probability. The overall patch-patch interaction energy between patches of different particles is given by
\begin{equation} \label{eq:epp}
    \beta\epsilon_{pp} = \beta \epsilon_{uu} (1-\Theta)^{2}+ 2\beta\epsilon_{uo}  \Theta (1-\Theta) + \beta\epsilon_{oo} \Theta^{2},
\end{equation}
where $\epsilon_{uu}$ is the interaction energy between two unoccupied patches, $\epsilon_{oo}$ is the interaction energy between two occupied patches,  and $\epsilon_{uo}$ is the contribution to the interactions between an occupied and an unoccupied patch. 

Previously \cite{roosen2014ion} we have simplified our model by assuming $\epsilon_{oo}=0=\epsilon_{uu}$. While we have found reasonable agreement between experiments of protein-salt mixtures and predictions of our model, we want to explore the richness of the model by systematically studying the influence of the various parameters, including $\epsilon_{oo}$ and $\epsilon_{uu}$, on the behaviour of the model. In addition, the more simplified model cannot distinguish between systems with different initial net charges (i.e, different proteins) since the initial repulsion is not considered. This extension could help to elucidate the observed trend in some experimental systems in which the initial repulsion due to the net charge of the proteins shifts the location of the critical point for phase separation to higher salt concentrations, as observed in the following reference \cite{maier2021human}. We expect that $\epsilon_{oo}$ and $\epsilon_{uu}$ account for repulsive contributions to the the overall interaction energy between patchy particles, as they account for interactions between patches with the same electrical charge. The influence of the repulsion parameters is examined within the liquid-vapour phase coexistence region, which is the key area of interest in the experimental systems we aim to describe using our model. By employing Eq.~\ref{eq:min}, we constrain the system to this region. However, adjusting the repulsion parameters could allow for the study of phase separation suppression. In this work, the suppression of the liquid-vapour phase transition induced by repulsion will not be investigated.

\subsection{Thermodynamic Model}
The fundamental thermodynamic behaviour of our patchy particle model is based on the Wertheim theory. \cite{wertheim1984fluids,wertheim1984fluidsII,wertheim1986fluids} In this framework, the 
free energy density is given by the sum of the free energy density of the reference system and a perturbation contribution due to bonding
between particles. In our work we use a hard-sphere fluid as reference system and employ the accurate thermodynamics based
on the Carnahan-Starling \cite{carnahan1969} equation of state. Clearly, it would be possible to replace the 
hard-sphere fluid by a more general reference system. To this end one would have to replace the chemical
potential and the pressure of the hard-sphere fluid by those of the reference system of choice. 
The contribution due to bonding between patches of different particles is given by  $f_{bond}$ that is the Helmholtz free energy per volume associated with bonding
\begin{equation}
    \beta f_{bond} = m \, \frac{\eta}{\nu_s} \left(\ln(1-p_b)+\frac{1}{2}p_b\right),
\end{equation}
where $\eta=4 \pi R^3 \rho/3 = \nu_s \rho$  is the packing fraction, $m$, as before, the number of patches per particle and $p_b$ the probability of a patch having formed a bond. Note that $p_b$ depends on the number density $\rho$ or alternatively $\eta$ and follows from the mass-action equation \cite{roosen2014ion}
\begin{equation}
\frac{p_b}{(1-p_b)^2} = m \frac{\eta}{\nu_s} \Delta,
\end{equation}
where $\Delta$ accounts for the spherical averaged interaction between bonds of patches 
 of two particles. Here we
follow Refs.~\cite{jackson1988,roosen2014ion} and assume a hard-sphere reference system and a short
ranged interaction between patches and obtain
\begin{equation} \label{eq:delta}
\Delta = 4 \pi g_{HS}(\sigma,\eta) K F, 
\end{equation}
where $g_{HS}(\sigma,\eta)$ is the contact value of the radial distribution function of the hard-sphere reference
system with diameter $\sigma$, $K$ is the bonding volume and $F$ is the angular average
of the Mayer-$f$ function of the patch-patch interaction
\begin{equation}
 F = \exp{(-\beta \epsilon_{pp})}-1.
\end{equation}
The total free energy density in our model is the sum of the ideal gas-, the hard-sphere- and the Wertheim contribution. The resulting chemical potential of the system consists therefore also of three terms, the ideal gas chemical potential, the excess chemical potential of the hard sphere reference system and the bonding term of the chemical potential. The total pressure of the system is the sum of the Carnahan and Starling contribution $\beta P_{cs}$ \cite{carnahan1969} plus the contribution given by the bonding term. While for the hard-sphere reference system the thermal energy $\beta = 1/(k_B T)$ is a trivial scaling factor, the bonding contributions are sensitive to it. For sufficiently low temperature the chemical potential and the pressure can develop a van-der-Waals loop that indicates the possibility for a liquid-vapour phase transition. Details on the expression for the chemical potential and pressure used are given in Ref.\cite{roosen2014ion}.
 
 \section{Results}
 In this section we study the effect of the different parameters on the liquid-vapour phase behaviour of the
 ion-activated patchy particles model. Within the model we can adjust and change each parameter independently and thereby obtain insights into the influence of each parameter on the phase boundaries.
\subsection{Interaction energy curves}
To this end we start with the interaction energy curves given by Eq.~(\ref{eq:epp}). In a previous study the
parameters $\epsilon_{oo}$ and $\epsilon_{uu}$ were set to zero and all the focus was put onto the
attraction between an unoccupied and an occupied patch via $\epsilon_{uo}$. It was found \cite{roosen2014ion}
that in this case the minimum of the energy curve is at $\Theta=1/2$ and that this minimum has to be
sufficiently deep (negative) for the system to phase separate. 

\begin{figure*}[!ht]
    \centering
    \includegraphics[scale=0.9]{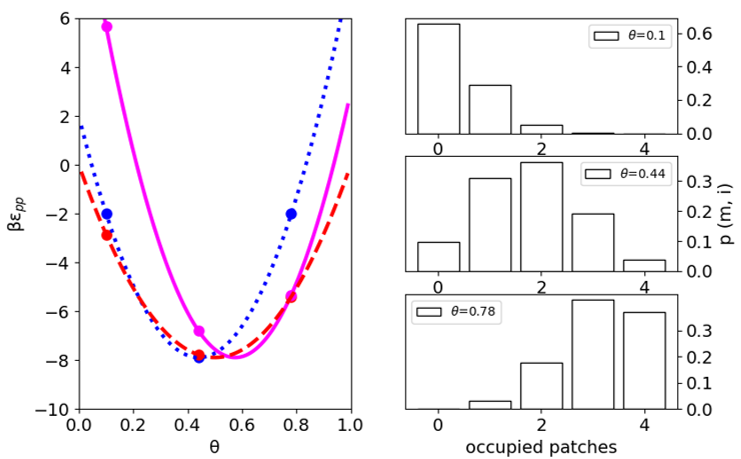}
    \caption{Interaction energies curves and corresponding histograms for three different choices of the interaction parameters. For the magenta curve the set of parameters used are $\beta\epsilon_{uu}$= 12, $\beta\epsilon_{oo}$= 3 and $\beta\epsilon_{uo}$= -22.60, for the blue curve $\beta\epsilon_{uu}$= 2, $\beta\epsilon_{oo}$= 8 and $\beta\epsilon_{uo}$= -20.42, and the red curve with  $\beta\epsilon_{uu}$= 0 and $\beta\epsilon_{oo}$= 0 and $\beta\epsilon_{uo}$= -15.77. Note that for three different values of $\Theta$, 0.1, 0.44, and 0.78, (marked as dots on the left figure) we show the corresponding histograms for the probabilities of finding particles with a different number of salt ions bound to it. While the histograms are determined by $\Theta$, the resulting pair interaction energy depends on the energy parameters $\beta\epsilon_{uu}$, $\beta\epsilon_{oo}$ and $\beta\epsilon_{uo}$.}
    \label{fig:energy}
\end{figure*}

 In order to display the influence of the parameters $\epsilon_{oo}$ and $\epsilon_{uu}$ on the overall interaction energy between particles $\epsilon_{pp}$, we compare it to the result of our previous model where those parameters were set to zero. In Fig.~\ref{fig:energy}, we plot the reference curve with no repulsion together with the curves obtained including the repulsive contributions. The value of the attractive interaction parameter 
 $\beta\epsilon_{uo}$ is fixed so that it produces curves with the same minimum $\beta\epsilon_{min}$. For a given set of parameters the condition is given by 
 \begin{equation}
     \beta\epsilon_{uo} = -\beta\epsilon_{min}-\sqrt{(\beta\epsilon_{uu}+\beta\epsilon_{min}) (\beta\epsilon_{oo} + \beta\epsilon_{min})}.
     \label{eq:min}
 \end{equation}
The results in Fig.~\ref{fig:energy} show that including the repulsive parameters generates an asymmetry in the interaction energies curves. In fact, we find that for the energy curve with no repulsion, the interaction energy curve is symmetric with the minimum located at $\Theta = 0.5$ (red curve), while for the other curves the location of the minimum is shifted to $\Theta\approx 0.44$ for $\beta\epsilon_{uu}$ = 2 , $\beta\epsilon_{oo}$= 8 and $\beta\epsilon_{uo}$ = -20.42 (blue curve) and $\Theta \approx 0.58$ for for $\beta\epsilon_{uu}$ = 12 , 
$\beta\epsilon_{oo}$= 3 and $\epsilon_{uo}$ = -22.60 (magenta curve). It is worth noting that changing those parameters does not lead to a change in the probability distribution $p(m, i)$ at a given value of $\Theta$, but the same histograms refer to different energy values in the interaction energy curves Eq.~(\ref{eq:epp}). This is an intriguing behaviour of our model that we want to underline, since different experimental systems, for example proteins, have been demonstrated to be very sensitive on the type of salt used. \cite{Matsarskaia_2018_PhysChemChemPhys, matsarskaia2016cation, senft2023effective,buchholz2023kinetics, surfaro2023alternative,maier2021human} In our model those effects can be described by the parameters $\beta\epsilon_{oo}$ and $\beta\epsilon_{uo}$. This in turn can change the resulting protein phase behaviour. In a sense our system, which is a mixture of particles and ions, can be interpreted as a multi-component system of proteins with a different number of salt ions bound. As shown in Fig.~\ref{fig:energy} a given value of $\Theta$, the binding probability of an ion to a patch together with the energy parameters determine the protein-protein interaction, Eq.~(\ref{eq:epp}), and the distribution of particles with a different number of ions bound to them, as shown in the histograms. The overall interaction energy is the key parameter to connect our ion-activated patchy particle model with the framework of the Wertheim theory. 

In the next section we will describe the effect of $\beta\epsilon_{uu}$, $\beta\epsilon_{oo}$ and $\beta\epsilon_{uo}$ on the phase diagram of our model.

\subsection{Effect of the interaction energy parameters on the liquid-vapour equilibrium}
\begin{figure*}[!ht]
    \centering
    \includegraphics[scale=0.71]{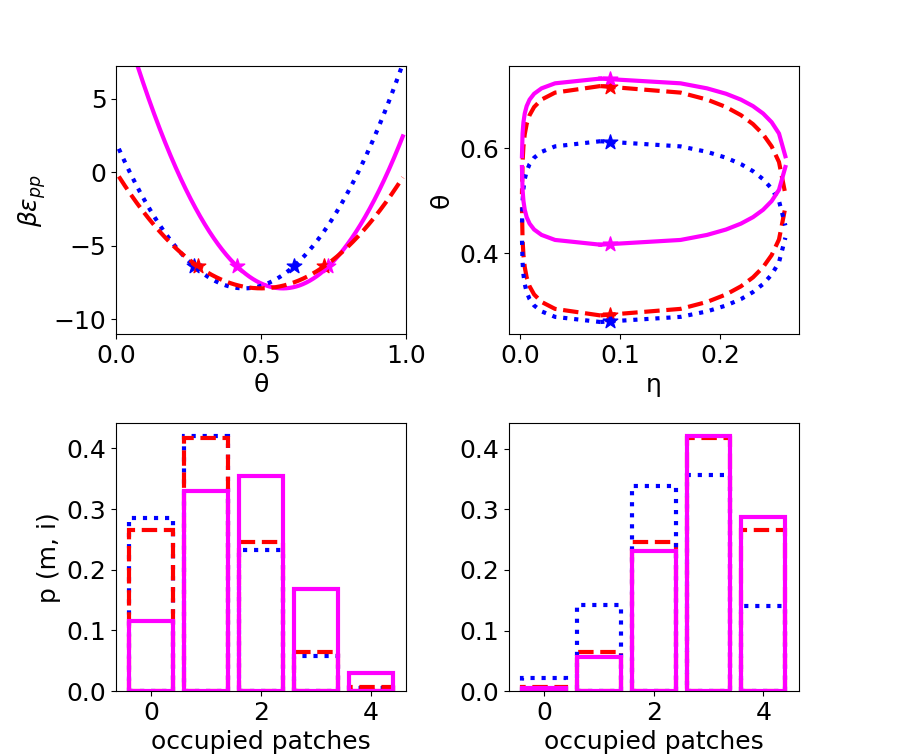}
    \caption{Energy curves and corresponding coexistence loops for three different choices of the interaction parameters
    $\beta \epsilon_{pp}$, as chosen in Fig. \ref{fig:energy}. Note that the energy parameters are chosen so that the
    minima of the energy curves have the same depth, leading to the same width in the coexistence loop, while their locations
    vary from case to case, which causes an up or down shift and a variation in the areas of the corresponding phase diagram. The histograms show the occupancy-distributions for the two critical points for the three choices
    of energy parameters. The histogram on the left-hand side corresponds to the lower critical point, while the histogram on the right-hand side corresponds to the upper critical point.}
    \label{fig:phase}
\end{figure*}

In order to test the effect of these parameters on the liquid-vapour coexistence regions we want to recall the condition for phase equilibrium between a low density phase with packing fraction $\eta_1$ and a high density phase at the same temperature with packing fraction $\eta_2$. The coexistence between phases implies the mechanical equilibrium $P(\eta_1) = P(\eta_2)$ and chemical equilibrium $\mu(\eta_1) = \mu(\eta_2)$. The Wertheim expressions for the pressure and the chemical potential do not allow for analytical solution and therefore the liquid-vapour equilibrium is evaluated numerically. In Fig.~\ref{fig:phase} we show the effect of the repulsion on the shape of the coexistence loop. 

As expected from the energy curves shown in Fig.~\ref{fig:energy}, changes in the repulsion contributions to the energy curves shift the coexistence loop up or down. Both the pressure and the chemical potential are, in fact, dependent on several parameters, including the number of patches $m$, the radius of the particles $R$, the packing fraction $\eta$, and the parameter $F$ and $K$, introduced in Eq.~(\ref{eq:delta}). In our calculations we keep $K$ fixed, unless mentioned otherwise. The quadratic form of the interaction energy, Eq.~(\ref{eq:epp}), implies that for a given interaction energy $\beta\epsilon_{pp}$ there are two different values of $\Theta$ or equivalently two different salt concentrations that give rise to the same protein-protein interaction energy, but with different compositions of occupied and unoccupied patches. Therefore, our model predicts a closed coexistence loop with two critical points \cite{roosen2014ion}. These points are highlighted by symbols in Fig.~\ref{fig:phase}, both in the interaction energy curves and in the phase diagram. It is worth noting that the effect of repulsion not only shifts the interaction energy curves $\beta \epsilon_{pp}$ to the left or to the right, but also its width. As a result, the area of the coexistence loop is also reduced compared to the one which does not include the repulsion parameters.
    
\begin{figure}[]
    \includegraphics[scale=0.8]{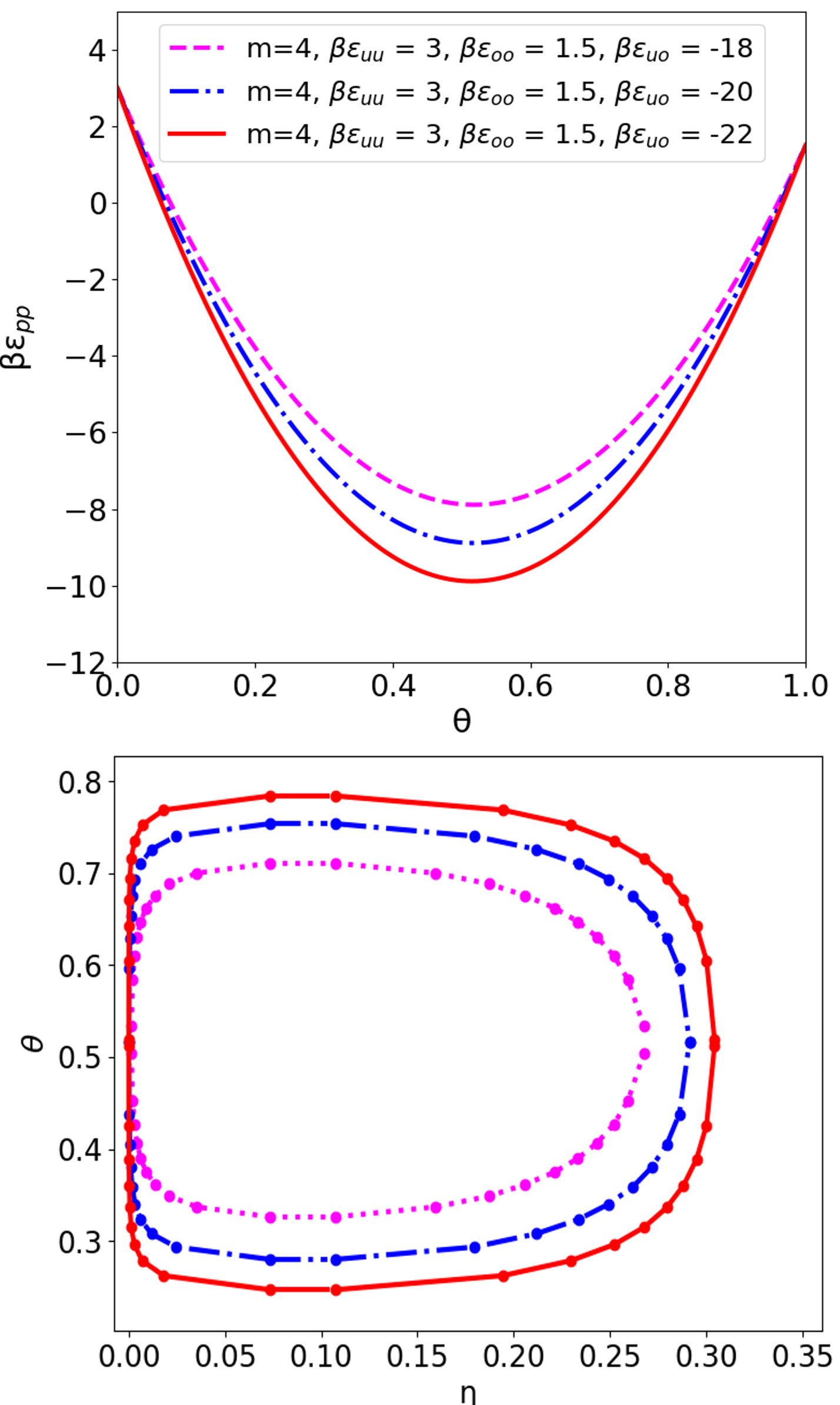}
    \caption{Interaction energy curves and coexistence loops for systems with the same repulsion parameters: Increasing the attractions increase the width of the interaction energy curves and therefore the size of the coexistence region.}
    \label{fig:attraction}
\end{figure}

The location of the critical point in the phase diagram in Fig.~\ref{fig:phase} is shifted along the probability or $\Theta$-axes, but not in interaction energy $\beta\epsilon_{pp}$ or packing fraction $\eta$. For this set of parameters ($m=4, R =1$), the critical value of the protein-protein interaction is approximately at $\beta\epsilon_{pp} \approx -6.38$. However, the distribution of occupied and unoccupied patches at the critical point is significantly different among the different curves, as depicted in the histograms. These differences might impact not only the shape of the coexistence loop and the liquid-vapour equilibrium but also on the more complex features of the phase diagram of patchy particles, for example, in the case of the formation of crystals or amorphous solids, as observed in protein-salt mixtures.\cite{sauter2014nonclassical,maier2021human} Consequently, appropriately configuring the repulsion introduces an asymmetry in the interaction energy curves, rendering the resulting behavior more accurate in describing protein solutions in the presence of trivalent salts, when compared with the simplified model that neglects the repulsion parameters. The size of the loop can be finely tuned by increasing or decreasing the depth of the interaction energy curve ($\beta \epsilon_{pp}$), while the position of the center along the y-axis ($\Theta$ in the phase diagram) can be tuned by changing the repulsion and shifting the interaction energy curve along the $\Theta$ axis on the ($\beta \epsilon_{pp},\Theta$) plane. Those effects are included in Fig.~\ref{fig:energy} and Fig.~\ref{fig:phase}. In Fig.~\ref{fig:attraction}, curves with the same values for the repulsion parameters $\beta \epsilon_{uu}$ and $\beta\epsilon_{oo}$ but, with different values of attraction parameter $\beta\epsilon_{uo}$ are shown. Again, the critical points are always located at the same packing fraction value but shifted in probability $\Theta$. The critical value for the packing fraction is equal to $\eta = 0.0898$ for the loops in both Fig.~\ref{fig:phase} and Fig.~\ref{fig:attraction}. That is consistent with previous works. \cite{liu2009vapor} However, the effect of increasing the attraction is not only changing the size of the loop but also, as shown in the previous case, changing the distribution of the particles at the critical point and at any given probability $\Theta$. 
In the next section we will discuss the effect of different thermodynamic parameters that are included in the Wertheim theory of the liquid-vapour equilibrium.
 \subsection{Effect of the thermodynamic parameters on the liquid-vapour equilibrium}
\begin{figure}[H]
    \centering
    \includegraphics[scale=0.5]{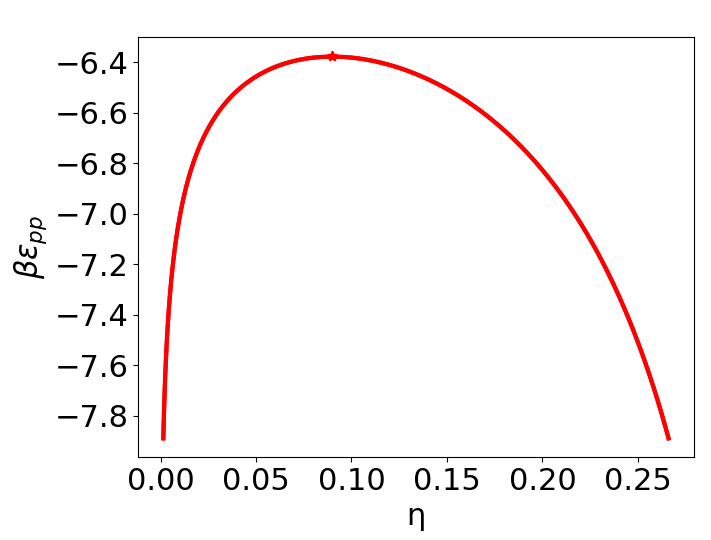}
    \caption{Wertheim liquid-vapour phase coexistence regions, the location of the critical point $\eta_c$ is determined by the initial choice of $m$ and $K$.}
    \label{fig:enter-label}
\end{figure}

Our model is based on the Wertheim theory of patchy particles that has a well defined set of parameters from which the thermodynamic quantities such as the pressure and chemical potential follow. For the liquid-vapour coexistence, the Wertheim theory gives a characteristic phase diagram shown in Fig.\ref{fig:enter-label}. This representation is unchanged for a given value of $K$ and $m$, as well as the position of $\eta_c$, and the coexistence regions are always between the critical value $\beta \epsilon_c \approx -6.38$ and the minimum $\beta \epsilon_{\text{min}} \approx -7.9$. In our model, the effective interactions are driven by $\beta \epsilon_{pp}$ as given by Eq.\ref{eq:epp}.Due to the quadratic form of the expression obtained with our model, there are now two different values of $\Theta$ that have the same interaction values of $\beta \epsilon_{pp}$ within the critical region for phase coexistence. Therefore, the resulting coexistence regions on the $(\Theta, \eta)$ plane are given by the combination of the $\beta \epsilon_{pp}$ curves given by Eq.\ref{eq:epp}, with the Wertheim representation on the $(\beta \epsilon_{pp}, \eta)$ phase diagram.

In this section, we want to explore the effect of changing these thermodynamic parameters employed within the Wertheim theory on the shape of the liquid-vapour equilibrium loop of our model.

The first parameter that we test is the number of patches $m$ on each proteins.
It is important to consider that the number of patches does not affect the overall shape of the interaction energy equation Eq.~(\ref{eq:epp}). 

\begin{figure}
    \centering
    \includegraphics[scale=0.61]{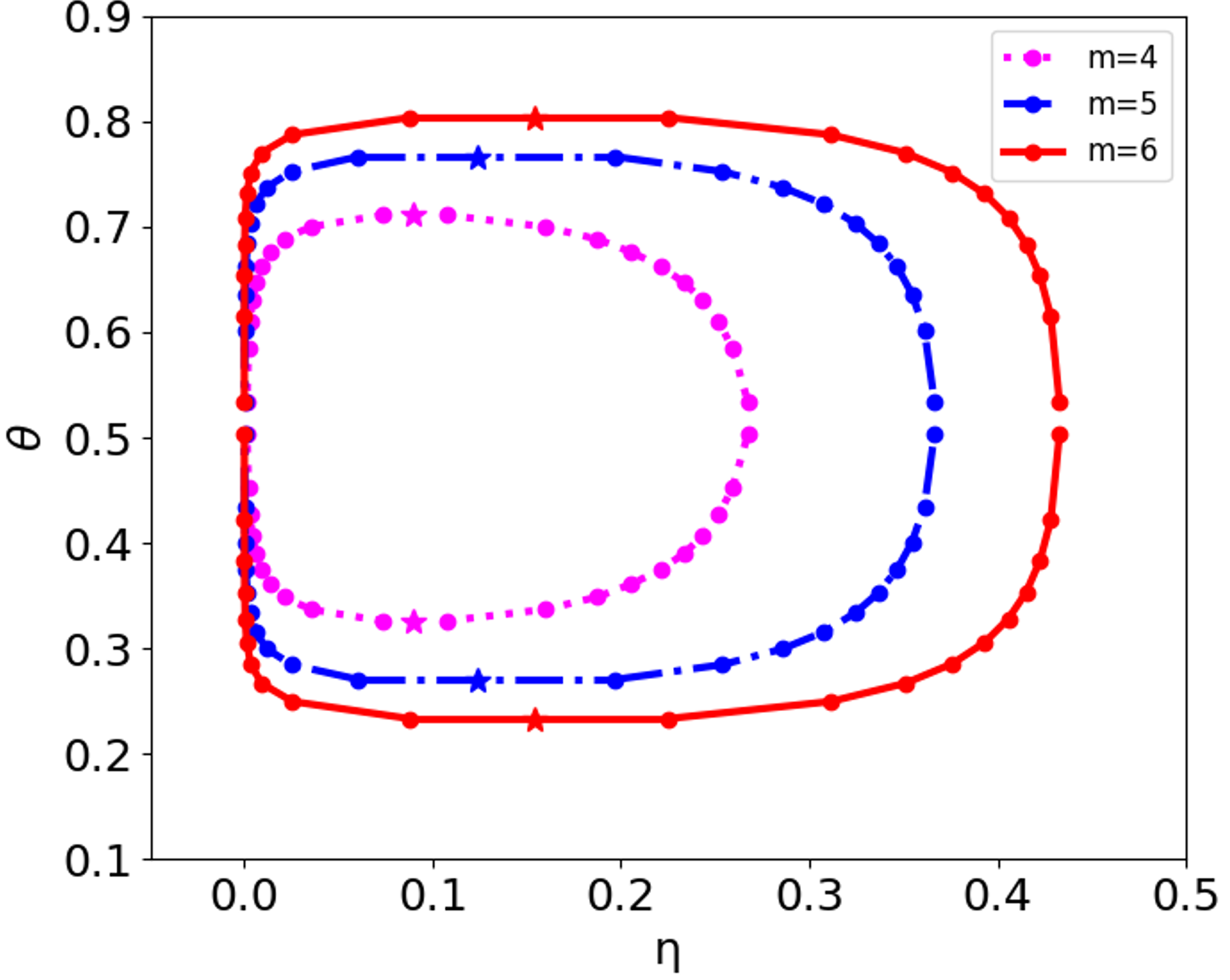}
    \caption{Effect of changing the number of patches on the patchy particles for interaction energy curves with $\beta\epsilon_{uu}$= 3, $\beta\epsilon_{oo}$= 1.5 and $\beta\epsilon_{uo}$= -18. Increasing the number of patches increases the interaction energy necessary to reach liquid-vapour phase separation, it leads to an increased size of the coexistence loop and changes the critical packing fraction. }
    \label{fig:patches}
\end{figure}

The dependence of the number of patches can be seen by the probability $p(m,i)$ of the occupied patches $\Theta^{i}$ and of unoccupied patches $(1-\Theta)^{m-i}$, given in Eq.~(\ref{eq:prob}). In Fig.~\ref{fig:patches} we show how the effect of changing the number of patches $m$ shifts the critical points for liquid-vapour phase separation to higher values of the packing fraction. This is consistent with previous works \cite{bianchi2006phase,liu2007vapor}. The shift in critical point, is due to the fact that $m$ enter directly in the thermodynamic framework of the Wertheim theory and is not a direct consequence of changing $p(m,i)$.

We want to emphasize that changing the number of patches does not only change the size of the coexistence loop but also the properties of the dense and diluted phase. In a system with only two patches, which does not phase-separate \cite{bianchi2006phase}, the possible geometry of clusters is limited to form linear chains that eventually might close into a ring. For a larger number of patches, different cluster morphologies are possible, which can form in the dense phase as well as in the gel. The effect of changing the number of patches does not affect only the network of the dense phase but also the properties of the solid, the crystalline and the gel structures. A detailed treatment of such effects in patchy particle systems can be found in Refs.~\cite{romano2012phase} and  \cite{giacometti2010effects}.
In particular, in Ref.~\cite{romano2012phase}, different types of crystalline phases are found for particles with valence of $m=3$ and compared with particles with $m=5$.

\begin{figure*}[]
    \centering
    \includegraphics[scale=0.59]{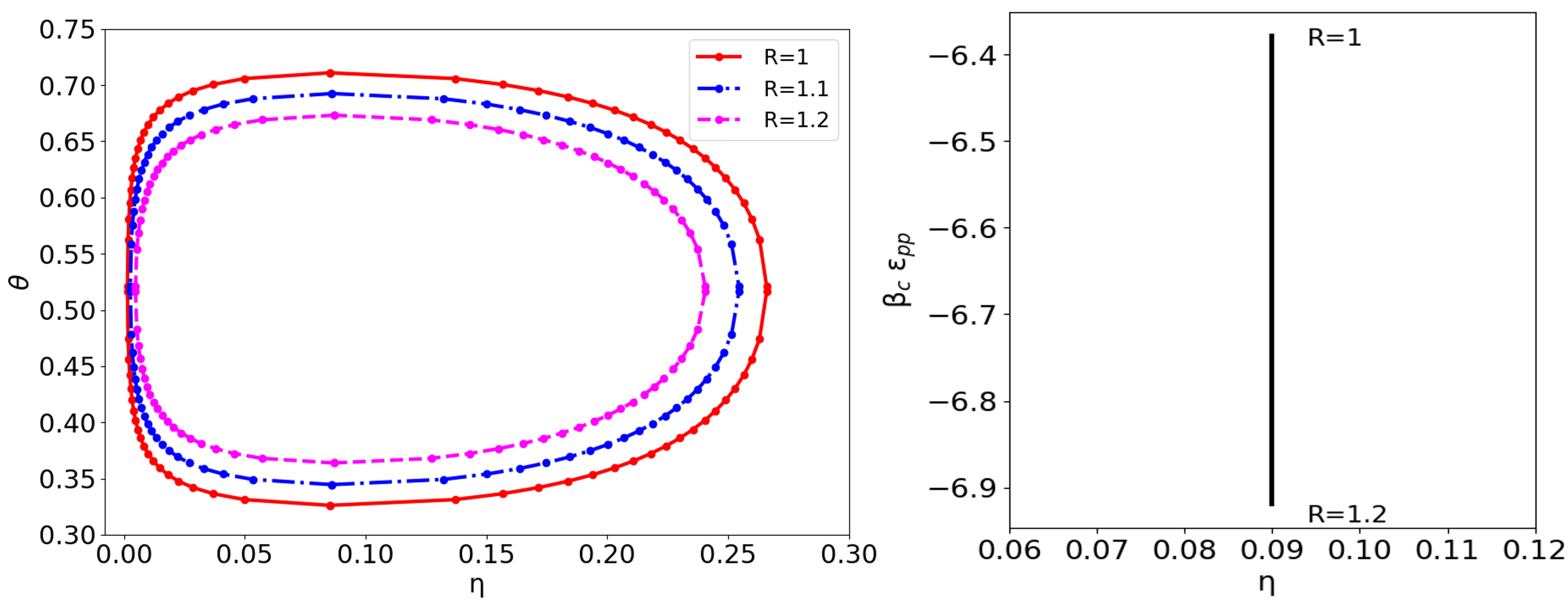}
    \caption{Effect of changing the radius of the particles for interaction energy curves with $\beta\epsilon_{uu}$= 3, $\beta\epsilon_{oo}$= 1.5 and $\beta\epsilon_{uo}$= -18. Increasing the radius of the particles decreases the interaction energy necessary to reach the critical point and reduces the size of the coexistence loop.}
    \label{fig:radius}
\end{figure*}

Another interesting effect is related to the other two parameters used in the Wertheim theory. The first one is the effect of the particle hard-sphere radius $R$ and the second one is related to the volume of the interaction $K$ referred as bonding volume. Changing the radius of the particle, the size of the loop change as well as the value of the interaction energy at the critical point $\beta_c\epsilon_{pp}$, but not the critical packing fraction $\eta_c$ which remains constant. An example of these effects are included in Fig.~\ref{fig:radius}. This effect is due to the fact that the ratio between the interaction volume parameter $K$ and the volume of the patchy particle increase as the radius decrease, resulting in stronger attractions. In other words, the fraction of the surface that is covered by patches increases. The effect of the radius might explain several phenomena in experimental systems. 

For example, in protein solutions, the interactions between solvent and macro-molecules can lead to a change of the hydrodynamic radius, depending on the protein-solvent interactions. We speculate, that within the framework of the Wertheim theory and ion-activated attractive patches model, it is possible to explain several effects arising from the protein-solvent interactions, without the need to explicitly treat the solvent, but indirectly infer the solvent contribution from the particle radius or from experimental systematic behaviour. In protein for example, the hydrodynamic radius is connected to protein-solvent interactions. Some differences in phase behavior were observed when replacing $H_2O$ with $D_2O$, potentially stemming from a more compact structure in heavy water, leading to slight variations in the hydrodynamic radius of proteins in the two solvents. \cite{braun2017strong, giubertoni2023d2o, cioni2002effect, sasisanker2004solvation} A strongly polar solvent, for example, increases  the hydrophobic effect and, depending on the amino-acid composition of the protein, might compact or relax the tertiary structure. In Fig.~\ref{fig:radius} we show that small changes in the radius of the particle have a significant impact on the concentration of the liquid-vapour equilibrium. For example an increase of 0.2 units, increases the resulting packing fraction in the low density phase of about 3.3 times and decreases the packing fraction of the dense phase of about 12.5$\%$ at the extreme points of the coexistence regions, while at central values the effects is less significant.
We can also induce the same effect reported in Fig.~\ref{fig:radius} by fixing the radius of the protein and increasing $K$. In this way we can tune the ratio between the interaction volume parameter and the volume of the patchy particle which is relevant to produce this effect.

 \section{Effect of the temperature}
\begin{figure}
    \centering
    \includegraphics[scale=0.65]{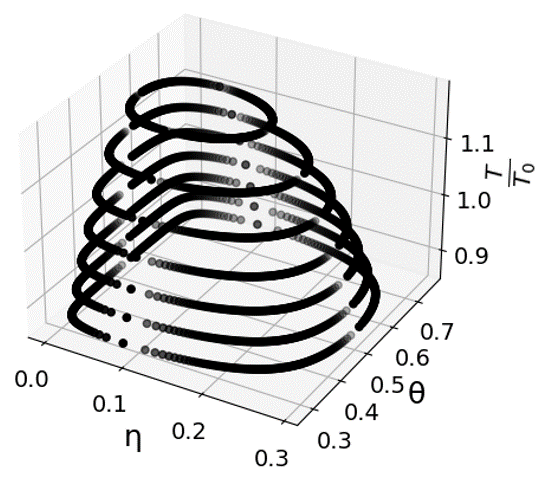}
    \caption{ Qualitative effect of changing the temperature on  the coexistence curves. The reference curve is at $T$ = $T_0$ and refer to $\beta\epsilon_{uu}$= 3,  $\beta\epsilon_{oo}$= 1.5 and $\beta\epsilon_{uo}$ = -18.}
    \label{fig:phase3d}
\end{figure}

The temperature is a fundamental thermodynamic parameter which has been kept constant in the considerations so far. However, as we increase temperature in our system the attraction between particles, induced by the formation of salt bridges between patches is weakened. If the temperature reaches a critical value $T_c$ liquid-vapour phase separation vanishes and above $T_c$ only the mixed fluid state is observed. Note that in this study we do not consider the solid phase. The full fluid phase diagram as a function of temperature $T$, the protein packing fraction $\eta$ and the binding probability $\Theta$ is shown in Fig.~\ref{fig:phase3d}. In the limited region of temperature explored in Fig.~\ref{fig:phase3d} we considered $\Theta$ independent of the temperature, since in this regard, the temperature is a scaling factor. As a result, there is a shift along the $\Theta$ axis if the temperature dependence is taken into account. However, the qualitative behaviour expected with a smaller coexistence loop at higher temperature and an increased size at lower values does not change.
In reality, the direct effect of temperature in experimental systems such as proteins can be far more complicated, involving not only a change of the interactions, but also strong conformational changes, that can make the spherical model assumed here not a good representation of the system. However, in the region of temperature where the globular structure of the protein is preserved, before denaturation, most of the effect due to the temperature are well reproduced within our model. It is worth noting that the temperature in the experimental system does not only affect the conformation of the proteins, but also changes the $K_w$ of the water equilibrium and the resulting hydration interactions between ions and water, protein and ions as well as the specific pKa values of the aminoacidic residues \cite{khechinashvili1995thermodynamics}. Therefore, some intrinsic variables in the experiment are not investigated due to limitation of the coarse grain approach.

\section{Conclusion}
In this work we have extended the ion-activated patchy particles model for protein-salt mixtures, taking into account the interaction between binding sites of proteins in more details than before. While in previous studies the main focus was on the attraction between proteins induced by a salt bridge that forms between an occupied and an unoccupied binding site, here we have also taken the repulsion between two unoccupied and between two occupied binding sites into account. The effect of these enriched interactions in our model produces a different and asymmetric ensemble of particles, consisting of proteins with no bound ion, proteins with one bound ions, etc. up to proteins with all binding sites filled, in the system. This rich multicomponent system, can be used to understand different phenomena, for example it can be used to rationalise different kind of phase transitions in protein-salt solutions, such as liquid-liquid phase separation.\cite{maier2021human,matsarskaia2016cation,surfaro2023alternative,buchholz2023kinetics,roosen2013interplay,roosen2014ion,senft2023effective,sauter2014nonclassical} 
We have shown that, although the model of ion-activated patchy particle model based on the Wertheim theory is rather simple, the parameters have important effects on the liquid-vapour phase separation loop. Our main results are:
\begin{itemize}
\item The effect of the repulsion between two occupied and between two unoccupied binding sites on proteins influences the height of the liquid-vapour loop by shifting the interaction energy curves and resulting in different patchy particles distribution as described by the histograms.
\item Increasing the attraction between an occupied and an unoccupied site increases the size of the coexistence loop as well as changes the ensemble distribution at the critical point.
\item The only sensitive parameter to change the critical packing fraction in the coexistence loop is the number $m$ of patches on the surface of the particles. Increasing the number of patches leads to a bigger size of the loop and shifts the critical interaction energy for phase separation to higher values. Increasing the number of patches also increase the number of possible components in the histograms.
\item Increasing the hard-sphere radius changes the ratio between the interaction volume parameter $K$ and the volume of the particle, decreasing the interaction energy necessary to reach the critical point and reducing the size of the phase-separation loop. 
\item Changing the bonding volume $K$ produces the same effect of changing the radius of the particles, but in the opposite direction since the bonding probability is going as $\approx K \nu^{-1}_s$. An increase of the bonding volume $K$ leads to an increased size of the loop due to the increased ratio between the volume of the interaction and the volume of the particle.
\end{itemize}
The aim of our model is to describe the phase behaviour of proteins in salt solutions. In the ion-activated patchy particle model the probability $\Theta$ to have an occupied patch on the surface of the protein is given by a Fermi-like distribution in the grand canonical ensemble Eq.~(\ref{eq:binding}). This is a function of the salt concentration in the reservoir, or equivalently its chemical potential $\mu_s$. The reservoir concentration of salt ions, $c_s^r$, is a quantity within the theoretical framework. However, the total salt concentration in the system, which is the quantity that can be controlled or measured in experiments, is directly connected with the salt concentration in the reservoir through \cite{roosen2014ion}
 \begin{equation}
    c_s = m \Theta \rho + c_s^r(\mu_s) (1-\eta(1+R_s/R)^3).
\end{equation}
The first term takes into account the ions bond on the surface of the patchy particle and the second term originates from the free ions in the solution, corrected for the volume excluded by the proteins, where $R_s$ is the radius of the salt ion. Using this relation, it is possible to access the concentration of ions on the surface of the patchy particles obtaining coexistence loops that can be compared to the experimental one, since they are a function of the salt concentration that it is also our experimental variable.
\begin{figure}
    \centering
    \includegraphics[scale=0.545]{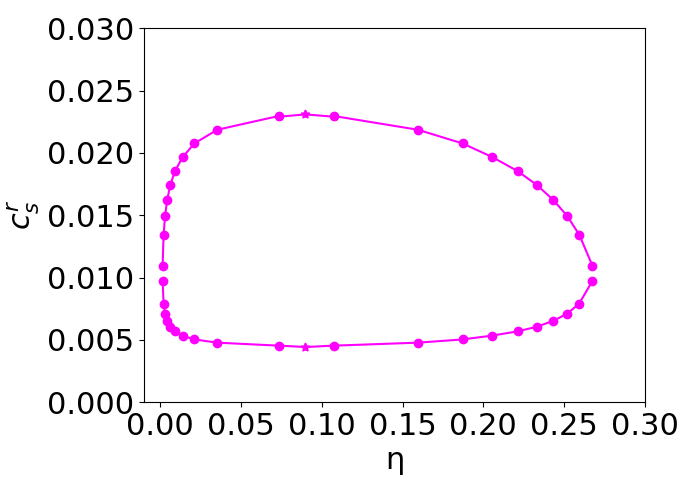}
   \caption{Liquid-vapour coexistence loop as a function of the salt concentration in the reservoir. The parameters considered are the following: $\beta \epsilon_{uu} = 3$, $\beta \epsilon_{oo} = 1.5$, $\beta \epsilon_{uo} = -18$, and the ion-protein binding energy $\beta \epsilon_{b} = -4.5$. The salt concentration is expressed in arbitrary units.}
    \label{fig:reservoir}
\end{figure}
In Fig. \ref{fig:reservoir} we show the phase diagram w.r.t the salt concentration in the reservoir for a given set of parameters.

It will be interesting to compare predictions of our enriched model to known experimental results. For example in Maier et al.\cite{maier2021human} BSA and HSA in presence of $CeCl_3$ display a shift in the location of liquid-vapour coexistence loop. This could arise because the initial net charge of BSA is more negative w.r.t HSA, it can be seen within our model as a more repulsive $\beta \epsilon_{uu}$. We expect that our model can be further enriched by considering our system as mixture of different kinds of particles as shown in Ref.\cite{braz2021phase, heidenreich2020designer}. Following this possible extension would be interesting to compare the phase diagram in Fig.\ref{fig:reservoir} with the phase diagram produced by considering the system as a multicomponent mixture. Furthermore, our model can be extended by allowing for binding sites with different binding energies. This will increase the complexity of the model and should be an important step towards understanding complex systems such as proteins in salt solutions.

\section{Acknowledgments}
The authors acknowledge the Deutsche Forschungsgemeinschaft (DFG) and the Bundesministerium für Bildung und Forschung (BMBF) for their generous funding, which made this research possible.

\bibliography{Reference.bib}

 \end{document}